\documentclass[aps,prl]{revtex4}

\usepackage{graphicx}
\usepackage{latexsym}
\usepackage{amsmath}

\def\br{\mathbf{r}}

\def\bk{\mathbf{k}}
\def\bG{\mathbf{G}}

\begin{document}

\title{Beyond time-dependent exact-exchange:\\
the need for long-range correlation}

\author{Fabien Bruneval$^{1,*}$, Francesco Sottile$^{1,2,*}$,
Valerio Olevano$^{1,3,*}$ and Lucia Reining$^{1,*}$}
\affiliation{$^1$ Laboratoire des Solides Irradi\'es UMR 7642,
CNRS-CEA/DSM, \'Ecole Polytechnique, F-91128 Palaiseau, France.
\\
$^2$ Donostia International Physics Center (DIPC),
20018 Donostia/San Sebastian, Spain.
\\ 
$^3$  Laboratoire d'\'Etudes des Propri\'et\'es \'Electroniques des Solides, 
UPR 11, CNRS, F-38042, Grenoble, France.
\\
$^*$ European Theoretical Spectroscopy Facility (ETSF).
}
\date{April 14, 2006. Published on J. Chem. Phys. {\bf 124}, 144113 (2006).}

\begin{abstract}
In the description of the interaction between electrons beyond the classical Hartree picture,
bare exchange often yields a leading contribution.
Here we discuss its effect on optical spectra of solids, comparing three different frameworks:
time-dependent Hartree-Fock, a recently introduced combined density-functional and Green's functions approach
applied to the bare exchange self-energy,
and time-dependent exact-exchange within time-dependent density-functional theory (TD-EXX).
We show that these three approximations give rise to identical excitonic effects in solids; these effects
are drastically overestimated for semiconductors.
They are partially compensated by the usual overestimation of the quasiparticle band gap within Hartree-Fock.
The physics that lacks in these approaches can be formulated as screening.
We show that the introduction of screening in TD-EXX indeed leads to a formulation
that is equivalent to previously proposed functionals derived from Many-Body Perturbation Theory.
It can be simulated by reducing the long-range part of the Coulomb interaction: this produces
absorption spectra of semiconductors in good agreement with experiment.
\end {abstract}

%\pacs{71.10.-w, 71.15.Qe, 71.35.-y, 78.20.Bh}

\maketitle

\section{Introduction}

Density Functional Theory (DFT) \cite{hk} is one of the most widely used approaches
for the calculation of material properties that are determined by the electronic ground state.
Since existing approximations for the in principle exact but in practice unknown exchange-correlation contribution are sometimes not sufficient to obtain the desired accuracy or even qualitative behavior,
the search for better functionals is a continuous effort.
The same is true concerning neutral electronic excitations
that are accessible observables of time-dependent DFT (TD-DFT) \cite{rungegross}.
The failure of most existing approximations for extended systems is in this case even more significant
than in the case of the ground state \cite{rmp_lucia}.
In both cases, it has been recognized that one may have to accept an additional complication of the functionals in order to get reliable results.

One way to go are orbital-dependent functionals.
A local Kohn-Sham (KS) \cite{ks} exchange-correlation (xc) potential $v_{xc}$ can be obtained
from a given non-local self-energy via the (linearized) Sham-Schl\"uter equation \cite{shamschluter_prl}.
This optimized effective potential (OEP) \cite{OEP}
contains much of the important physics of the underlying self-energy.
The most prominent approximation along this path is the so-called ``exact exchange'' (EXX) \cite{gorling_new}.
The EXX potential $v_\text{EXX}$ is obtained as the local counterpart for the non-local Fock operator,
which is usually referred to as the ``exchange'' term.
This potential is of course completely different from the original Fock operator:
the OEP is local and constructed with KS wavefunctions,
whereas the exchange operator is non-local and constructed with Hartree-Fock (HF) wavefunctions.
It has been found empirically that EXX eigenvalue band gaps are closer to experimental quasiparticle band gaps
than are local-density approximation (LDA) or HF ones.
The agreement is very good for simple semiconductors \cite{exx_aulbur}.
However, an increasing underestimation for materials with wider band gap has been noticed
\cite{exx_ar}.
Although EXX contains important effects (it is devoid of self-interaction \cite{gorling_new}, as HF is),
it is not meant to simulate the effects of Hartree-Fock.
In particular, there is no simple interpretation for the KS eigenvalue band gap,
no equivalent to the Koopmans theorem.
Since KS band gaps are {\it a priori} not meant to reproduce experiment,
the band gap agreement may arise to a large extent from error cancellations (though still be useful).

Instead, in the case of electronic excitations, the situation is different.
Time-Dependent Hartree-Fock (TD-HF) can be understood as an approximation to Many-Body Perturbation Theory (MBPT),
obtained from the latter by neglecting correlation.
TD-EXX, on the other side, is a straightforward  approximation of TD-DFT.
Both MBPT and  TD-DFT should in principle yield the same (correct) dynamical polarizability,
that can be measured e.g. by optical absorption. 

HF and, more recently, EXX have been used quite frequently in calculations of real solids \cite{dovesi,rinke},
sometimes augmented with approximate correlation functionals, like LDA \cite{exx_gs}.
Much fewer examples exist instead for their time-dependent counterpart;
TD-HF has been carried out for large band gap materials \cite{hanke_sham_diamond},
and, to our knowledge, only the absorption spectrum of silicon has been calculated within TD-EXX
\cite{gorling_tdexx_prl}.
Intuitively, one would of course not choose TD-HF to calculate, say, the absorption of silicon,
since the strong screening in this material can be expected to drastically influence the electron-hole interaction.
On the other hand, with EXX KS band gaps comparing so favorably to experiment with respect to their HF counterparts,
one may hope for a similar improvement when going from TD-HF to TD-EXX.
In fact, it was suggested in Ref.~\onlinecite{gorling_tdexx_prl}
that the TD-EXX absorption spectrum of bulk silicon favorably compares to experiment.
It is therefore worthwhile to elucidate the links between the two approaches
and discuss differences in the various ingredients as well as in the expected and calculated results.
This is the main aim of the present work.

For the sake of clarity and completeness, we add to this comparison a third method,
that is to some extent intermediate.
This recently introduced approach, called $\rho/G$ in the following,
is in fact a combination of MBPT and the density-functional concept \cite{fabien}.
As will be discussed below in the exchange-only approximation, it is situated on the same level as HF concerning the quasiparticle band gap,
but close to TD-EXX concerning the electron-hole attraction.

The three methods used in the present work are different for some aspects, but tightly linked for
others. 
For a better understanding, we summarize in Table~\ref{tab:schema} the differences and similarities of the three frameworks.
The meaning of the notations and the acronyms will be made clear along this paper.

Our mathematical comparison and numerical results (obtained for the example of bulk silicon) clearly show
that in the case of solids all three methods are equivalent.
This implies that TD-EXX in its present form is not suitable for the description of absorption spectra
of semiconductors.
We discuss hence the need for screening of the exchange interaction,
and, more precisely, for the screening of the long-range components of it.
By introducing the missing terms, we make the link with recently introduced and successfully used functionals
derived from Many-Body Perturbation Theory \cite{fabien,roro,adragna,francesco_prl,marini_boundexc},
and we discuss which are the most crucially needed corrections.

%%%%%%%%%%%%%%%%%%%%%%%%%%%%%%%%%%%%%%%%%%%%%%%%%%%%%%%%%%%%%%%%%%%%%%%%%%%%%%%%%%%%%%%%%%%%%%%
%\begin{turnpage}

\begin{table}[h]
\caption{
Schematic overview of the different approaches and corresponding observables, approximations and
acronyms. Quantities are specified for the exchange-only case.
\label{tab:schema}
}
{\scriptsize

\begin{tabular}{lcccc}
\hline\hline
 & & & & \\
                   & \multicolumn{2}{c}{(TD-)DFT}       &     $\rho/G$             &   MBPT               \\
                   &  &                 &                          &                      \\
%Electronic Density
%     & \multicolumn{2}{c}{$\rho^\text{EXX}(\br t) = \rho^\text{HF}(\br t)$}
%     & $\rho^\text{HF}(\br t)$
%     & $\rho^\text{HF}(\br t)$ \\
Green's function
     & \multicolumn{2}{c}{$G_{EXX}(\br_1 t_1, \br_2 t_2)$}
     & $G_\text{HF}(\br_1 t_1, \br_2 t_2)$
     & $G_\text{HF}(\br_1 t_1, \br_2 t_2)$ \\
%Central variable   &  \multicolumn{2}{c}{$\rho(\br t)$}   
%                   & $\rho(\br t)$, $G(\br_1 t_1,\br_2 t_2)$
%		   & $G(\br_1 t_1, \br_2 t_2)$             \\
 & & & & \\
 & \multicolumn{2}{c}{$\overbrace{\phantom{aaaaaaaaaaaaaaaaaaaaaaaaaaaaaaaaaaaaaaaaaaa}}$} & & \\
                   & Lin. Sham-Schl\"uter     &  Sham-Schl\"uter     &  & \\
Potential          &  $v_\text{EXX}$          & generally not used 
                   &   $\Sigma_x$             & $\Sigma_x$           \\
Eigenvalues        &  $\epsilon_\text{EXX}$   & generally not used
                   &   $\epsilon_\text{HF}$   & $\epsilon_\text{HF}$      \\
 & & & & \\
\hline
 & & & & \\
\parbox[c]{3.5cm}{Non-interacting \\
response function used \\ in polarizability equation}
                   &  \multicolumn{2}{c}{$\chi_0^\text{EXX}$} 
                   &  $\chi_0^\text{HF}$  
		   &  ${}^4\chi_0^\text{HF}$    \\
 & & & & \\
 & \multicolumn{2}{c}{$\overbrace{\phantom{aaaaaaaaaaaaaaaaaaaaaaaaaaaaaaaaaaaaaaaaaaa}}$} & & \\
                   & Lin. Sham-Schl\"uter     &  Sham-Schl\"uter     &  & \\
 & & & & \\
Kernel of the Dyson
equation            &  $K^\text{TD-EXX}$ 
                    &  $K^\text{TD-DFT}$  &  $K^{\rho/G}$  &  ${}^4K(1,2;1',2')$  \\
consisting of: & & & & \\
Variation of Hartree potential
                   & $v$
                   & $v$
                   & $v$
                   & $\delta(1,1') \delta(2,2') v(1,2)$   \\
Quasiparticle shift
                   &  $f^{(1),\text{lin}}$
		      ($= F_\text{EXX}^B$ of Ref.~\onlinecite{gorling_tdexx_prl})
		   &  $f^{(1)}  =\chi_0^{\text{EXX} -1} - \chi_0^{\text{HF} -1}$
                   &   ---             &    ---        \\
Electron-hole interaction
 &  $f^{(2),\text{lin}}$  ($= F_\text{EXX}^A$ of Ref.~\onlinecite{gorling_tdexx_prl})
 &  $f^{(2)}=\chi_0^{\text{HF} -1} G_\text{HF}G_\text{HF}vG_\text{HF}G_\text{HF} \chi_0^{\text{HF} -1}$
 &  $f^{(2)}$
 &  \parbox[c]{3cm}{$1/i \delta \Sigma_x / \delta G_\text{HF} = {}^4v =$ \\
 $= \delta(1,2) \delta(1',2') v(1,1') $}             \\
 & & & & \\
Name of the approaches
                   &  TD-EXX
                   &  \textit{Non-lin.} TD-EXX
                   &  \parbox[c]{2cm}{Exchange-only approx. of $\rho/G$}
                   &  TD-HF  \\
For solids, see papers
                   &  Kim and G\"orling \cite{gorling_tdexx_prl}
                   &  Bruneval \textit{et al} \cite{fabien}
                   &  Bruneval \textit{et al} \cite{fabien}
                   &  Hanke and Sham \cite{hanke_sham_diamond}  \\
 & & & & \\
\hline\hline
\end{tabular}
}

\end{table}
%\end{turnpage}

%%%%%%%%%%%%%%%%%%%%%%%%%%%%%%%%%%%%%%%%%%%%%%%%%%%%%%%%%%%%%%%%%%%%%%%%%%%%%%%%%%%%%%%%%%%%%%%

\section{Electron-hole interaction in the quasiparticle and in the density-functional framework}

It is useful to first recapitulate the main features of the various approaches that will be compared.
For the sake of clarity, we simplify the problem 
and do not distinguish between KS and HF single-particle wavefunctions in the following.
In the numerical examples, we use LDA wavefunctions throughout,
and furthermore, all the band structures (LDA, HF, EXX) differ solely by a rigid shift
of the band gap.
These assumptions are very reasonable for bulk silicon \cite{hybertsenlouie_prb,fabien_thesis} 
(of course, the situation would be rather different in finite systems \cite{baerends}).
Atomic units are used throughout the present work. We often employ the widely spread many-body 
short-hand notation 
$1= (\br_1,t_1)$ and omit to specify spin explicitly.

The neutral excitations of materials are described by the polarizability
or density-density linear response function $\chi({\bf r},{\bf r}',t-t')$.
This quantity expresses the linear response of the electronic density to
variations of an external potential $\delta U_{ext}(\br',t')$:
\begin{equation}
\chi({\bf r},{\bf r}', t - t') =  \left.
\frac{\delta \rho(\br,t)}
     {\delta U_{ext}(\br',t')} \right|_{U_{ext} = 0}.
\end{equation}
Beside the response of independent particles $\chi_0({\bf r},{\bf r}',t-t')$,
the full density-density response function further contains contributions stemming
from the self-consistently induced potentials.
Independent particle and interacting particle response functions are linked
via a Dyson-like polarizability equation, symbolically:
\begin{equation}
\label{eq:chi-gen}
\chi = \chi_0
  + \chi_0 \kappa \chi ,
\end{equation}
where the kernel $\kappa$ first of all is due to a self-consistently  induced Hartree potential
$v_H^{ind}({\bf r},t) = \int d{\bf r}'dt'\left[ \delta v_H({\bf r},t)/\delta \rho({\bf r}',t')\right] \delta \rho({\bf r}',t')$.
The induced Hartree potential contributes hence to the kernel $\kappa$ with the simple term
$\left[ \delta v_H({\bf r},t)/\delta \rho({\bf r}',t')\right]=v({\bf r}-{\bf r}')\delta(t-t')$.

A similar variation has to be added for the exchange-correlation potential. In the TD-DFT framework, this leads to the exchange-correlation kernel $f_{xc}({\bf r},{\bf r}',t-t') = \left[ \delta v_{xc}({\bf r},t)/\delta \rho({\bf r}',t')\right]$,
and Eq.~(\ref{eq:chi-gen}) is an integral equation for $\chi({\bf r},{\bf r}', t - t')$:
\begin{equation}
\chi(1,2)=\chi_{0}^\text{KS}(1,2)
 + \int d3 d4  ~
 \chi_{0}^\text{KS}(1,3) K^\text{TD-DFT}(3,4) \chi(4,2)   .
\label{eq:2POINT_INTRO}
\end{equation}
Here, $K^\text{TD-DFT}(3,4) = v(3,4) + f_{xc}(3,4)$.
$v(1,2)=v(\br_1-\br_2) \delta(t_1-t_2)$ stands for the instantaneous bare Coulomb interaction
and $\chi_{0}^\text{KS}$ is the Kohn-Sham independent-particle response function.

In the MBPT framework, 
 the equivalent variation has to be calculated for the non-local exchange-correlation self-energy $\Sigma$.
As a consequence, it is not possible to obtain straightforwardly a closed Dyson-like equation
for the two-point density response function. This remains also true in the HF approximation,
where $\Sigma$ reduces to the -- still non-local -- Fock operator $\Sigma_x$. 
Moreover, the exchange term has a simple dependence on the one-particle
Green's function $G(\br_1 t, \br_2 t^+)$
[not so simple on the density $\rho ({\bf r})$].
Therefore, as in the general case, the derivation of a closed equation for the response function of the Hartree-Fock system
leads to the introduction of four-point polarizabilities $^4\chi_0$ and $^4\chi$: \cite{rmp_lucia}
\begin{equation}
{^{4}}\chi(1,2;1',2')={^{4}}\chi_{0}(1,2;1',2')
 + \int d3 d4 d5 d6  ~
 {^{4}}\chi_{0}(1,3;1',4) {^{4}}K(3,6;4,5) {^{4}}\chi(5,2;6,2')   .
\label{eq:4POINT_INTRO}
\end{equation}
[Contracting indices to $^4\chi(1,2;1,2)$ gives back the usual $\chi(1,2)$.]
This equation is the Bethe-Salpeter equation \cite{hanke_sham,rmp_lucia}.

For the Hartree-Fock case, the kernel ${}^4K$ is
\footnote{Due to the $\delta$ functions acting on time arguments,
the equation can be written in frequency domain,
as a function of a single frequency.}
\begin{eqnarray}
{^{4}}K(3,6;4,5)=
2 \delta(3,4)\delta(5,6) v(3,5)
- \delta(3,6)\delta(4,5) v(3,4)  .
\label{eq:4POINT_KERNEL}
\end{eqnarray}
The first $v$ of the kernel stems from the variation of the Hartree potential
(with a factor of 2 for singlet excitons),
and the second $v$, from the variation $[\delta \Sigma_x/\delta G]$ of the Fock exchange operator $\Sigma_x$.
${^{4}}\chi_{0}$, the 4-point independent-quasiparticle polarizability,
is constructed here using HF eigenvalues, ${^{4}}\chi_{0}^\text{HF}$ (see last column of Table~\ref{tab:schema}).
Note that beyond bare exchange, in most applications of the Bethe-Salpeter equation
a statically screened Coulomb interaction $W$ replaces $v$ in the second term \cite{rmp_lucia}.
This term  gives rise to a direct screened electron-hole attraction instead of the unscreened one
in the case of HF.

Recently, Bruneval {\it et al.} \cite{fabien} introduced a general formulation
that avoids the solution of the four-point Bethe-Salpeter equation.
This approach (named $\rho/G$ here) is based on the use of the density-functional concept
within Hedin's equation of MBPT \cite{hedin}.
Of interest here, it yields an explicit formula for the {\it two-point} polarizability $\chi$
that contains all exchange-correlation effects
via the density-variation of the self-energy, $\delta\Sigma/\delta\rho$,
and via the explicit use of an independent-quasiparticle (QP) polarizability.
The approach is between TD-DFT [because it leads to a two-point polarizability equation
like Eq.~(\ref{eq:2POINT_INTRO})]
and MBPT (because the QP, instead of the KS, independent-particle response appears).
Using the $GW$ approximation for $\Sigma$ \cite{hedin} and some further straightforward approximations,
this approach leads to previously proposed equation for $\chi$
that yields spectra in excellent agreement with the spectra calculated from Bethe-Salpeter
equation, and consequently, with experimental spectra \cite{roro,adragna,francesco_prl,marini_boundexc}.
This equation is similar to Eq.~(\ref{eq:4POINT_INTRO}), but recast into a two-point form:
\begin{equation}
\chi(1,2)=\chi_{0}(1,2) + \int d3d4 ~ \chi_{0}(1,3) K^{\rho/G}(3,4) \chi(4,2) ,
\label{eq:2POINT_rhoG}
\end{equation}
where the two-point kernel $K^{\rho/G}$ reads
\begin{eqnarray}
\label{eq:2K_a}
K^{\rho/G}(1,2) & := &   v(1,2) + f^{(2)}(1,2)  \\
  & = &  v(1,2) 
 - i \int d3 d4 d5 ~ \chi_0^{-1}(13)G(34)G(53)
                    \frac{\delta\Sigma(45)}{\delta \rho(2)}  .
\end{eqnarray}

For the present work, it is interesting to choose the Fock operator
as an approximation for the self-energy in $\delta\Sigma/\delta\rho$,
and consistently, to use the HF independent-particle response function $\chi_0^\text{HF}$
for $\chi_0$ in the polarizability equation (second column of Table~\ref{tab:schema}).
The two-point kernel $K^{\rho/G}$
of the Hartree-Fock problem becomes, following Ref.~\onlinecite{fabien},
\begin{equation}
K^{\rho/G}(1,2)
 \simeq   v(1,2)
   + \int d3 d4 d5 d6 ~ \chi_0^{\text{HF} -1}(13)G_\text{HF}(34)G_\text{HF}(53)
          v(45)G_\text{HF}(46)G_\text{HF}(65)\chi_0^{\text{HF} -1}(62)  ,
\label{eq:2K}
\end{equation}
where $G_\text{HF}(1,2)$ is the HF Green's function
and hence $\Sigma_x = iG_\text{HF}v$ the Fock operator.
Here, as in Ref.~\onlinecite{fabien}, we have used the approximation
$\delta G / \delta \rho = - G(\delta G^{-1} / \delta \rho ) G\simeq G \chi_0^{-1} G$.

\begin{figure}[t]
\includegraphics[width=0.9\columnwidth]{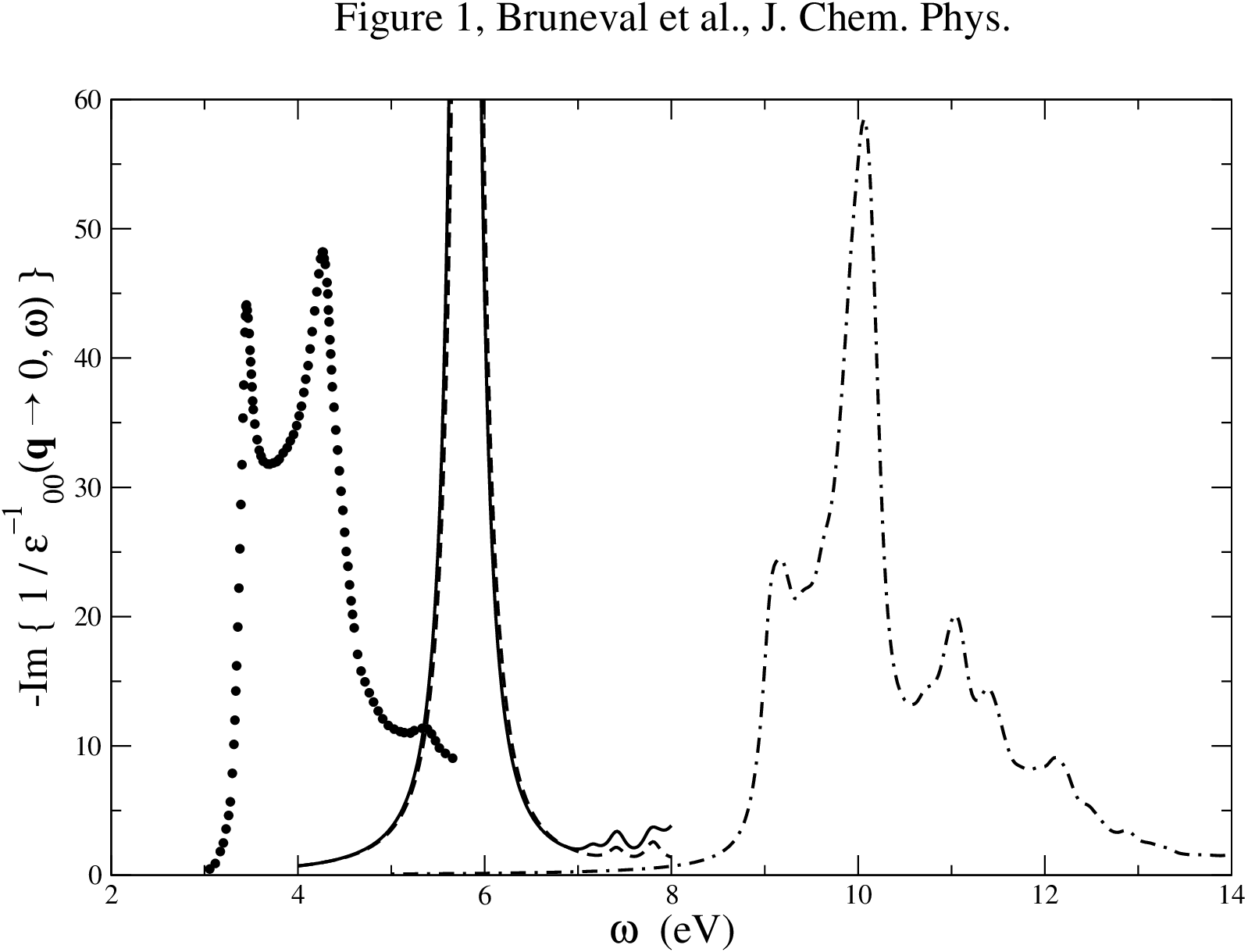}
\caption{Imaginary part of the macroscopic dielectric
function of bulk silicon.
Continuous curve: TD-HF result; 
dashed line: $\rho/G$ result (see text) from Ref.~\onlinecite{fabien};
dot-dashed curve: independent-HF-QP result (see text);
circles: experiment \cite{cardona_si}.
}
\label{fig:si_hf}
\end{figure}

One may wonder whether Eq.~(\ref{eq:2POINT_rhoG}) using this approximate $\rho/G$ kernel is indeed able to reproduce TD-HF.
Fig.~\ref{fig:si_hf} compares  the TD-HF (continuous curve) and $\rho/G$ (dashed curve) calculated
\footnote{
All calculations are made using norm-conserving pseudopotentials and plane-waves.
For the optical spectrum of bulk silicon, we employed
the three top valence bands and the three bottom conduction bands.
The Brillouin zone has been sampled by a set of 512 $\bk$-point.
The dimension of the grid is $8\times8\times8$ and it is
centered at (1/64, 2/64, 3/64) in reciprocal lattice units.
All matrices are expanded in a basis set of 169 plane-waves.
An imaginary part of 0.10\,eV has been used in the denominator of the response functions.
}
absorption spectra for bulk silicon. Of course, the results are far from any experiment:
the HF direct band gap of silicon is 8.92~eV
[see the independent-HF-QP result, dot-dashed curve obtained from $\text{Im}(\chi_0)$], more than twice the experimental QP direct band gap of 3.40~eV \cite{si_bandgap}.
Also the electron-hole attraction is drastically overestimated due to the absence of screening,
and a strongly bound exciton is formed inside the HF-QP band gap.
Finally, QP and excitonic errors cancel to a large extent;
the absorption spectrum falls in an energy region
that is closer to the experimental one (circles) \cite{cardona_si},
but the lineshape is of course completely wrong.
This is to be expected and is not the point here.
Instead, it is noteworthy to point out that the TD-HF [Eqs.~(\ref{eq:4POINT_INTRO}) and~(\ref{eq:4POINT_KERNEL})]
and $\rho/G$ [Eqs.~(\ref{eq:2POINT_rhoG}) and~(\ref{eq:2K})] results are almost indistinguishable,
as it was the case in previous findings
when correlation beyond Hartree-Fock was taken into account \cite{francesco_prl} and the effective interaction was therefore much weaker. 
This means that {\it the kernel $f^{(2)}$ in Eq.~(\ref{eq:2K}) simulates well the TD-HF bare electron-hole attraction}.

Let us now come to a fully TD-DFT formulation of the problem
(see the first column of Table~\ref{tab:schema}).
This can also easily be written starting from the equations of Ref.~\onlinecite{fabien}:
indeed, that work showed that a differentiation of the time-dependent Sham-Schl\"uter condition
\cite{vLeeuwen}
(that the TD-DFT and the MBPT time-dependent densities correspond)
with respect to the density yields the TD-DFT polarizability equation,
\begin{equation}
\chi(1,2) =\chi_{0}^\text{KS}(1,2) + \int d2 d3 ~ \chi_{0}^\text{KS}(1,3) K^\text{TD-DFT}(3,4) \chi(4,2) ,
\label{eq:2POINT_INTRO_DFT}
\end{equation}
in which now the KS independent-particle response $\chi_{0}^\text{KS}$ appears,
and the corresponding kernel reads $K^\text{TD-DFT} = K^{\rho/G} + (\chi_{0}^{\text{KS}-1}- \chi_0^{-1})$.
In other words, the TD-DFT kernel has, with respect to the $\rho/G$ one,
an additional contribution $f^{(1)} :=\chi_0^{\text{KS} -1}-\chi_{0}^{-1}$.
If inserted in a Dyson-like polarizability equation, $f^{(1)}$ transforms
the KS independent-particle response $\chi_{0}^\text{KS}$ into the corresponding 
QP independent-particle response $\chi_{0}$.
It essentially opens the band gap from the KS to the QP one \cite{fabien,pankratov}.

When applied to the exchange-only case,
one obtains hence $K^\text{TD-DFT}=v+f^{(1)} + f^{(2)}$
where $f^{(1)}$ has the role to open the band gap to the HF one,
and the bare electron-hole attraction $f^{(2)}$ is the kernel of Eq.~(\ref{eq:2K_a})
[approximated e.g. by Eq.~(\ref{eq:2K})]. 
We call this approach the ``non-linearized TD-EXX approach'',
as opposed to the standard TD-EXX approach that will be discussed in the next section.

It is quite obvious to see that Eq.~(\ref{eq:2POINT_INTRO_DFT}) and Eq.~(\ref{eq:2POINT_rhoG})
yield identical results, as we have also confirmed numerically (without displaying the result here).
This leads to an important conclusion of this section, namely:
{\it  non-linearized TD-EXX reproduces TD-HF}.
The band gap difference between EXX and HF is cancelled
in the optical spectrum by the contribution $f^{(1)}$ to the kernel,
whereas the effect of the electron-hole attractions $\delta \Sigma_x / \delta G$
and $f^{(2)}$ are extremely close.

\section{Time-dependent Hartree-Fock and time-dependent exact exchange}

Let us now make the link to what is usually called ``TD-EXX''.
The starting point is the  linearized Sham-Schl\"uter equation, \cite{vLeeuwen}
\begin{equation}
i \int d1 ~ v_\text{EXX}(1)\chi_0^\text{EXX}(1,2)
= 
\int d1 d3 \int d4  ~
 G_\text{EXX}(1,3) \Sigma_x(3,4) G_\text{EXX}(4,2) ,
\label{eq:vexx}
\end{equation}
where only EXX KS quantities are used to build response functions and the Fock operator $\Sigma_x = iG_\text{EXX}v$
(the solution of the static version of this equation would yield the static OEP potential $v_\text{EXX}$). 
The functional derivative with respect to the density of $v_\text{EXX}$
has a contribution that stems from the derivative of $\Sigma_x$
(this is explicitly shown in appendix~\ref{appendix:lin_ss}),
\begin{equation}
\label{eq:f2lin}
f^{(2),\text{lin}}(1,2) = \int d3 d4 d5 d6 ~
 \chi_0^{\text{EXX} -1}(1,3) G_\text{EXX}(3,4) G_\text{EXX}(5,3)
   v(4,5) G_\text{EXX}(4,6) G_\text{EXX}(6,5) \chi_0^{\text{EXX} -1}(6,2) .
\end{equation}
This term is very similar to $f^{(2)}$ in Eq.~(\ref{eq:2K}).
Since we do not distinguish single-particle wavefunctions,
the only difference lies in the eigenvalues used to build all the Green's functions, and $\chi_0$.

\begin{figure}[t]
\includegraphics[width=0.9\columnwidth]{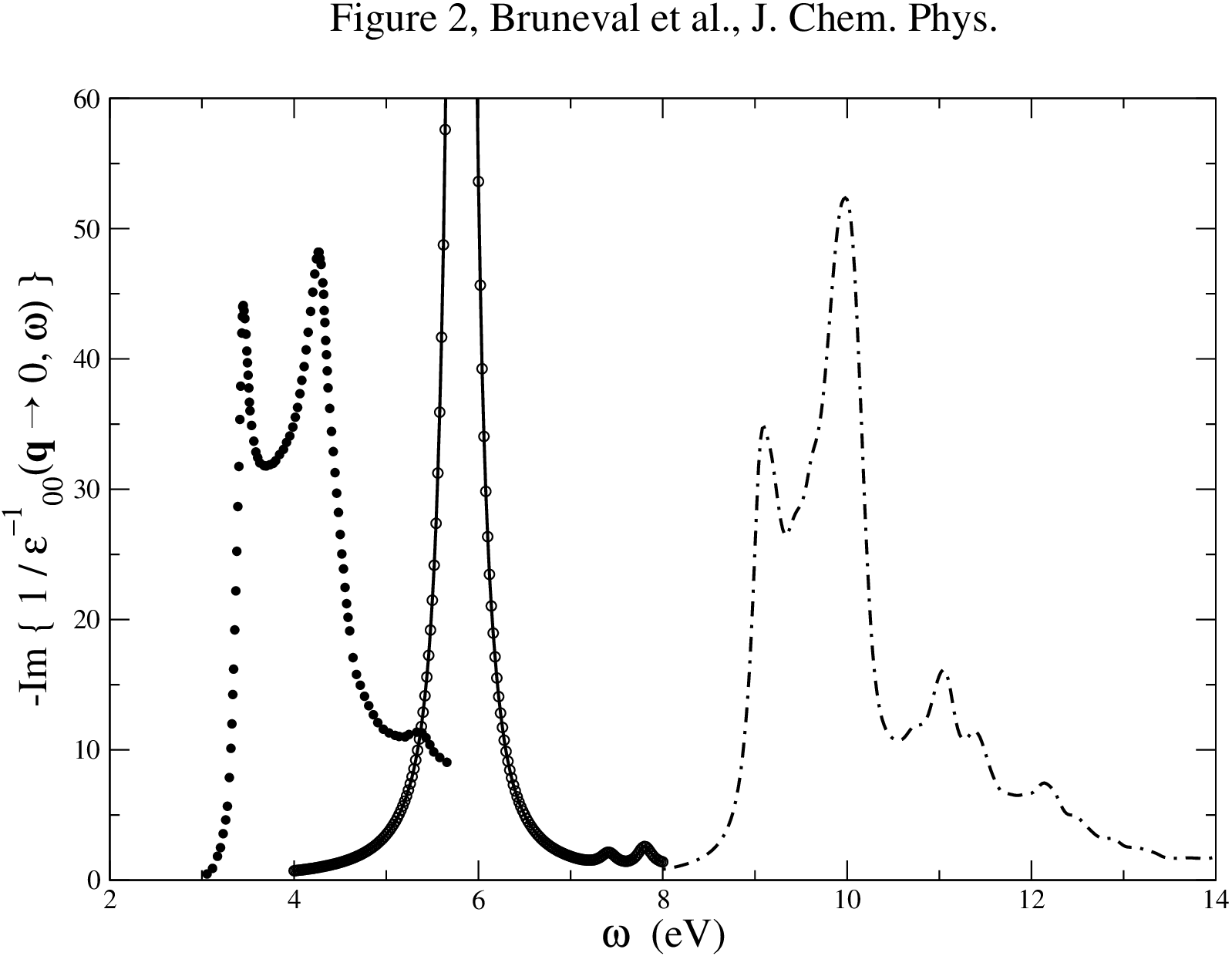}
\caption{
Imaginary part of the macroscopic dielectric
function of bulk silicon.
Continuous curve: $\rho/G$ result (see text) using $f^{(2)}$;
open circles: $\rho/G$ result using $f^{(2),\text{lin}}$ [this corresponds to TD-EXX (see text)];
dot-dashed curve: using $f^{(2),\text{lin}}$ and the modified Coulomb interaction of Ref.~\onlinecite{gorling_tdexx_prl}.
}
\label{fig:si-rho-G}
\end{figure}

Actually, it turns out that this difference is not significant.
Fig.~\ref{fig:si-rho-G} shows a comparison of the $\rho/G$ spectrum calculated
using $f^{(2)}$ and $f^{(2),\text{lin}}$,
respectively (continuous line and open circles).
The reason for this perfect agreement is a strong cancellation between energy denominators in the inverse response functions
and the two Green's functions in terms of the type $GG\chi_0^{-1}$.  

It is now important to notice that $f^{(2),\text{lin}}$ is nothing else but the electron-hole attraction term of TD-EXX.
This term corresponds precisely to the terms $H_X^1$ and $H_X^2$ in Ref.~\onlinecite{gorling_tdexx_prb}
(see appendix~\ref{appendix:kimgorling} for a detailed derivation).
Hence, {\it we find a strongly overbound exciton  from the TD-EXX electron-hole attraction}.

The rest of the terms that stem from the derivative of Eq.~(\ref{eq:vexx}) is the linearized version of  $f^{(1)}$.
It reads (see appendix~\ref{appendix:lin_ss})
\begin{multline}
\label{eq:f1lin}
f^{(1),\text{lin}}(1,2) =
  \int d3 d4 d5 d6 ~ \chi_0^{\text{EXX} -1}(1,3) G_\text{EXX}(3,6) G_\text{EXX}(6,4)
  \left[ \Sigma_x(4,5) -  \delta(4,5) v_\text{EXX}(4) \right] 
  G_\text{EXX}(5,3) \chi_0^{\text{EXX} -1}(6,2) \\
+ \int d3 d4 d5 d6 ~ 
   \chi_0^{\text{EXX} -1}(1,3) G_\text{EXX}(6,3) G_\text{EXX}(3,4)
 \left[ \Sigma_x(4,5) - \delta(4,5) v_\text{EXX}(4) \right]
  G_\text{EXX}(5,6) \chi_0^{\text{EXX} -1}(6,2)  .
\end{multline}
As shown in appendix~\ref{appendix:kimgorling},
this expression (again assuming that KS and HF wavefunctions are equal) corresponds
to the terms $H_X^3$ and $H_X^4$ of Ref.~\onlinecite{gorling_tdexx_prb}. 

$f^{(1)}$ has the difficult task to shift the whole independent-particle spectrum above the HF band gap,
and it turns out that it is more delicate to linearize this contribution than $f^{(2)}$.
It is clear that the linearization of $f^{(1)}$ cannot, by miracle, cancel
the overestimate of the exciton binding in $f^{(2)}$:
in the best case (met for few transitions), $f^{(1),\text{lin}}$ is a good approximation to $f^{(1)}$
and rigidly shifts the whole spectra conserving the shape (and therefore the bound exciton);
otherwise, $f^{(1),\text{lin}}$ is numerically instable and gives rise to scattered spectra \cite{T1}.
Therefore, in the following calculations, we always use the non-linearized version $f^{(1)}$
(instead of $f^{(1),\text{lin}}$).
For the same reason, it is not astonishing that Kim and G\"orling have found
a ``collapse'' of the silicon absorption spectrum \cite{gorling_tdexx_prl}.
The authors have solved the problem by cutting off the long-range (small-$q$) part
of the Coulomb interaction.
In fact, this procedure leads to drastic changes in the spectrum:
we have repeated their calculation by introducing the same cutoff in $f^{(2)}$.
Now, instead of the strongly bound exciton we find a lineshape in good agreement with experiment,
as can be seen by the dot-dashed curve in Fig.~\ref{fig:si-rho-G}.
($f^{(1)}$ has not been modified; therefore the spectrum stays in the HF energy region.)
In the same way,
the reduction of the long-range Coulomb interaction in $f^{(1)}$
translates into a shift of the spectrum towards the experimental position.

This result may seem rather {\it ad hoc}.
However, it can be understood and used to improve the approach,
as we will discuss in the following section.

\section{Correlation contributions}

\begin{figure}[t]
\includegraphics[width=0.9\columnwidth]{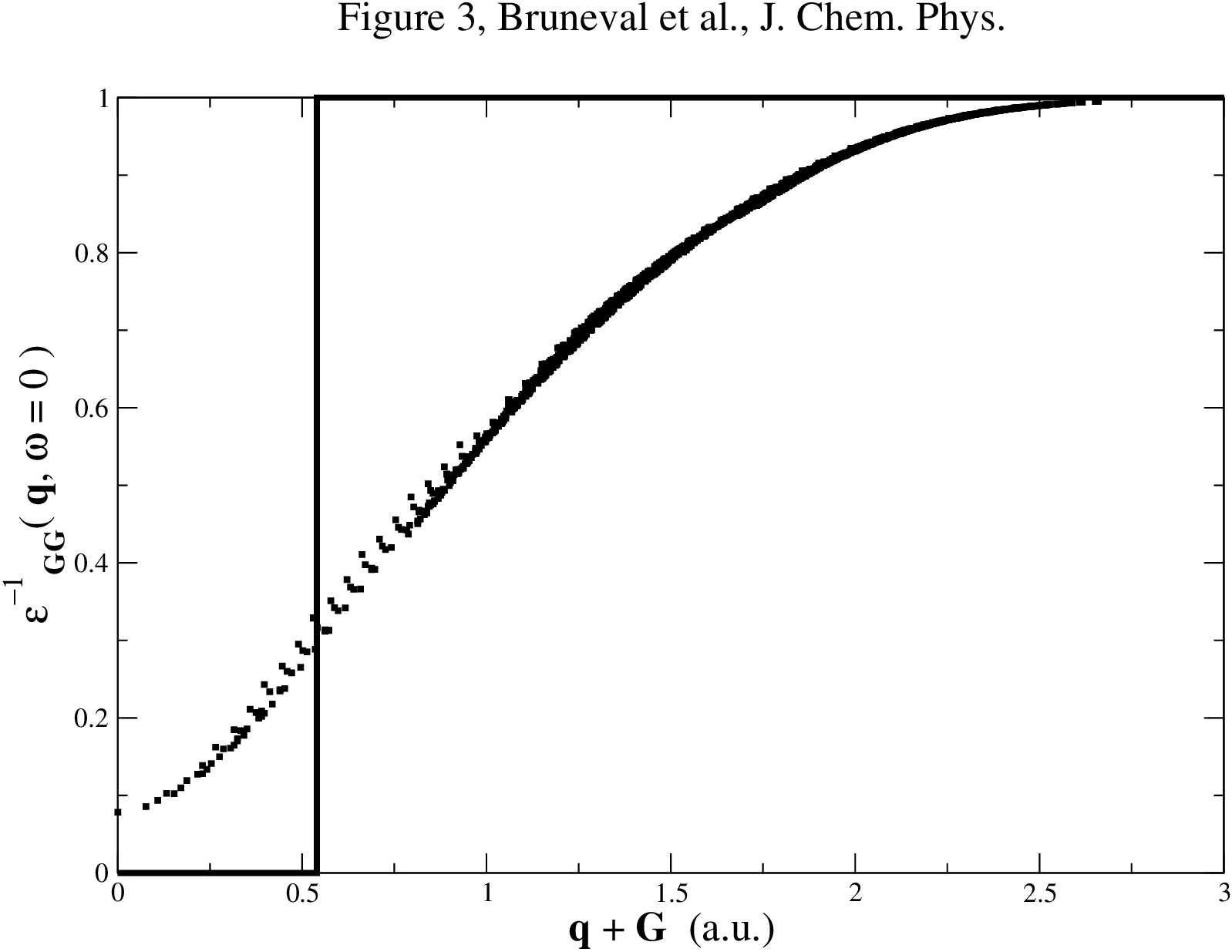}
\caption{
Diagonal of the static inverse  RPA dielectric function of bulk silicon.
The vertical line denotes the border of the Brillouin zone.
}
\label{fig:epsm1q}
\end{figure}

Fig.~\ref{fig:epsm1q} shows the diagonal of the static inverse dielectric matrix $\epsilon_{\bG,\bG}^{-1}(\bf q)$ of bulk silicon 
as a function of $|\bf q+\bf G|$, calculated in the Random Phase Approximation (RPA).
The vertical line denotes the border of the Brillouin zone,
up to which Kim and G\"orling \cite{gorling_tdexx_prl} have chosen to set the Coulomb interaction to zero.
The step function that one obtains in this way can be seen as a first reasonable approximation for the full screening curve.
In other words, the modified Coulomb interaction is an approximation to the screened Coulomb interaction $W$:
it compensates for the lack of  correlation.
Hence, the impressingly good result of Kim and G\"orling and in Fig.~\ref{fig:si-rho-G} can be
explained:
the new, screened $f^{(2)}$ is just an approximation to the electron-hole attraction term derived
in Refs.~\onlinecite{roro,adragna} from the Bethe-Salpeter equation,
which it reproduces in the same way as the unscreened version reproduces TD-HF.
The same applies in principle to $f^{(1)}$.

It should be pointed out that
the good results obtained from the more sophisticated approaches \cite{roro,adragna}
like $\rho/G$ \cite{fabien} rely in practice
on a number of approximations that are commonly made in the Bethe-Salpeter approach from which they are derived.
In particular, QP eigenvalues are calculated within the $GW$ approximation (including dynamical effects),
whereas $W$ for the electron-hole screening is taken static.
Although these are much less crude approximations than the cutoff used above for TD-EXX,
the search for a perfectly rigorous,
but still efficiently working approach is not yet completed.

At present, however, one may be with no doubts satisfied with the precision
and reliability of the screened approaches \cite{roro,adragna,francesco_prl,marini_boundexc}.
Nevertheless, it is interesting to investigate the role of correlation further, since the cutoff approach of Kim and G\"orling gives precious hints: the screening of the long-range (small-$q$) part of $v$ is seen to have drastic effects. This is consistent with other studies of the long-range contribution of the exchange-correlation kernel in bulk materials
(see e.g. Refs.~\onlinecite{alpha,BKBLS}). 

\begin{figure}[t]
\includegraphics[width=0.9\columnwidth]{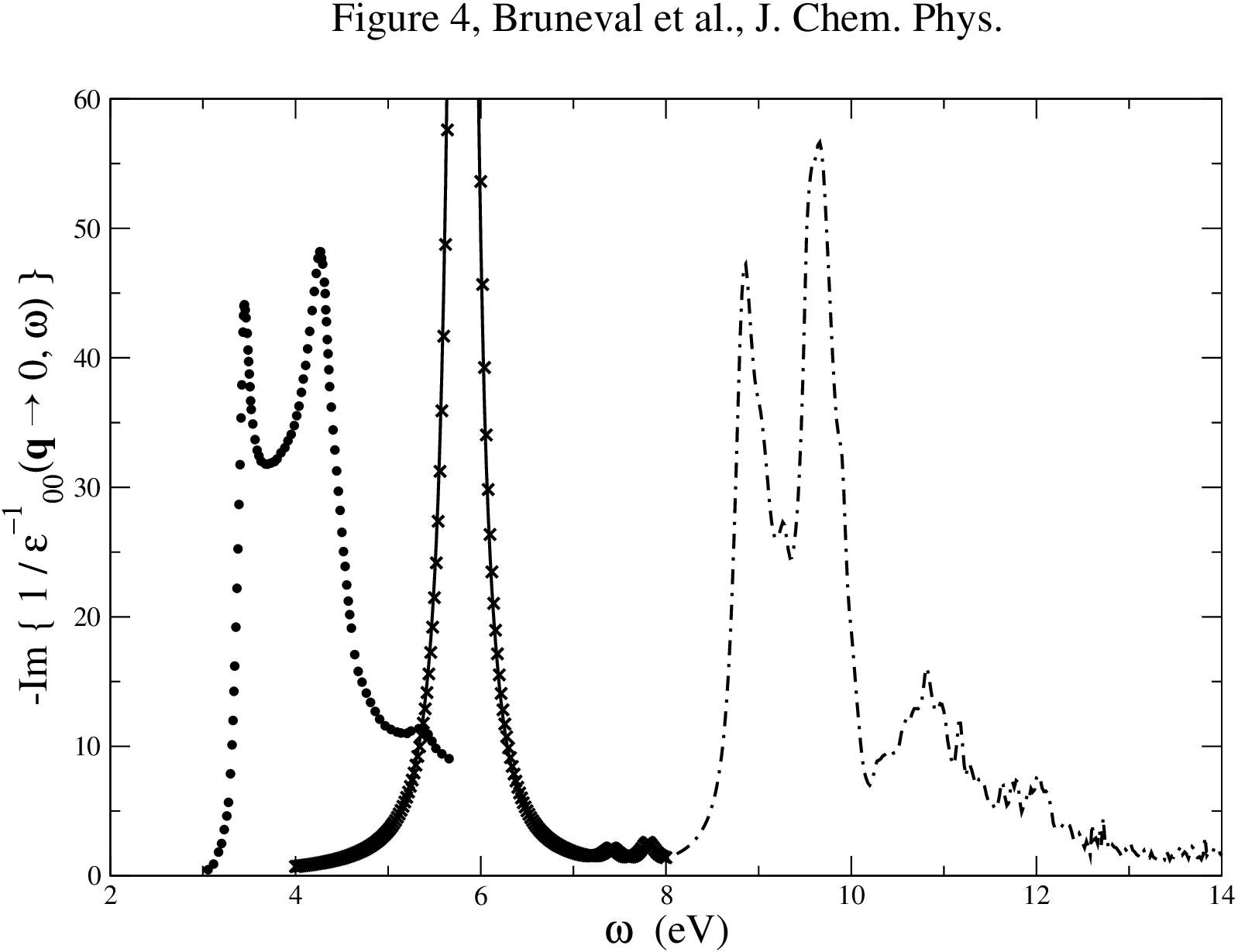}
\caption{
Imaginary part of the macroscopic dielectric
function of bulk silicon.
Continuous curve: non-linearized TD-(EXX+cLDA) result;
stars: non-linearized TD-EXX result;
dot-dashed curve: using $f^{(2),\text{lin}}$ and the modified Coulomb interaction $v/6$.
}
\label{fig:si-cLDA}
\end{figure}

Without considering these findings, one might hope to introduce correlation in another, more standard way, namely by adding LDA correlation to EXX as it is quite frequently done for the ground state potential \cite{exx_gs}.
Fig.~\ref{fig:si-cLDA} shows the result of a TD-(EXX+cLDA) calculation (continuous curve) as compared to the TD-EXX result (stars):
the LDA correlation does not give rise to any visible changes.
(Note that in both cases $f^{(1),\text{lin}}$ has been replaced by the corresponding $f^{(1)}$,
for clarity. In other words, only the effect of correlation in $f^{(2)}$ is tested.)

The adiabatic LDA kernel is in fact short ranged, and cannot suppress the overbound exciton stemming from long-range contributions. Instead, as can be seen from the cutoff result, any model $\epsilon^{-1}$ that reasonably screens the long range contributions can do the job;
for illustration, we also show in Fig. 4 the result obtained replacing $v$
by $v/\langle \epsilon_0 \rangle$ in $f^{(2),lin}$,
where $\langle \epsilon_0 \rangle$ is taken to be an average dielectric constant of 6 for silicon. In spite of the extreme simplicity of the treatment of correlation, the result is again satisfactory
(note that, as in Fig.~\ref{fig:si-rho-G},
the unscreened $f^{(1)}$ is used; hence, the spectrum results too high in energy).
This leaves the hope that, starting from some screened version of TD-EXX
and along the lines of Refs.~\onlinecite{roro,adragna,francesco_prl,marini_boundexc,pankratov_prl}, 
it is possible to find approximations to TDDFT,
less rigorous than TD-EXX,
but exempt from its severe shortcomings 
concerning the description of bulk materials, and that
the method is still numerically advantageous 
with respect to the solution of the Bethe-Salpeter equation.
Hybrid functionals like the ones discussed in Ref.~\onlinecite{scuseria} may be seen
as a possibility in this context.

\section{Conclusions}

A time-dependent OEP procedure constructs TD-DFT kernels
that yield the same time-dependent density as a given approximation
to the self-energy, via a time-dependent Sham-Schl\"uter equation.
Apart from a linearization, this relation is exact.
It is therefore not surprising that a careful time-dependent OEP
calculation reproduces the solution of the Bethe-Salpeter equation, within
the corresponding approximation.
The present work verified this agreement for the case of the Hartree-Fock approximation
to the self-energy.
In particular, the {\it non-linearized} and the usual linearized TD-EXX are shown to reproduce 
the TD-HF calculation and consequently, fail crudely to describe 
absorption spectra of semiconductors, because the electron-hole interaction
is largely overestimated there.
We show that TD-EXX gives rise to a huge bound exciton for bulk silicon. The linearization does not cure this shortcoming.
The similarity between TD-HF and TD-EXX can also be noticed in the evaluation
of vertical excitation energies of finite systems
\cite{td-exx_molecules}.

Whatever the approach used, Bethe-Salpeter equation, $\rho/G$ method,
or TD-DFT, the inclusion of the screening of the exchange operator is
evidenced as crucial to give a proper account for the electron-hole interaction. In particular,
the long-range components of the exchange kernel have to be reduced.
Furthermore, we demonstrate that the details of the screening do not
matter much: two very different and crude models (a cutoff as used by Kim and G\"orling
or a uniformly reduced Coulomb interaction $v/6$)
allow us to equally produce realistic absorption spectra of silicon.

The hope of describing optical absorption spectra of semiconductors
within TD-DFT is well founded.
However, in the framework of unscreened methods, like TD-EXX is,
there is no chance to get something else than the disastrous TD-HF results:
one has to go beyond TD-EXX.
A rigorous OEP method based on ``exact screened exchange'' may do the job.
Fortunately, the inclusion of screening within rather simple approximations,
e.g. using an empirically screened Coulomb interaction,
seems to be already sufficient.

\begin{acknowledgments}
We thank Andrea Cucca for his help.
We are grateful for discussions with R. Del Sole and A. G\"orling,
and for computer time from IDRIS (project 544).
This work has been supported by the EU's 6$^\text{th}$ Framework Programme
through the NANOQUANTA Network of Excellence
(NMP4-CT-2004-500198).
\end{acknowledgments}

%%%%%%%%%%%%%%%%%%%%%%%%%%%%%%%%%%%%%%%%%%%%%%%%%%%%%%%%%%%%%%%%%%%%%%%%%%%%%%%%%%%%%%%%%%%%%%
\appendix

\section{Derivation of the TD-DFT kernels from 
the linearized TD-Sham Schl\"uter equation}
\label{appendix:lin_ss}

The present appendix provides the derivation of the linearized TD-EXX kernels
from the linearized TD-Sham-Schl\"uter equation.
The linearized TD-Sham-Schl\"uter equation reads
\begin{equation}
\label{eq:lin_ss}
\int d3 ~ G_\text{EXX}(1,3) v_\text{EXX}(3) G_\text{EXX}(3,1) 
 = \int d3 d4 ~ G_\text{EXX}(1,3) \Sigma_x(3,4) G_\text{EXX}(4,1)  .
\end{equation}
When Eq.~(\ref{eq:lin_ss}) is differentiated with respect to the TD density $\rho(2)$,
we get
\begin{multline}
\label{eq:lin_ss_diff}
\int d3 ~ G_\text{EXX}(1,3) \frac{\delta v_\text{EXX}(3)}{\delta \rho(2)} G_\text{EXX}(3,1) 
 =  \int d3 d4 ~ G_\text{EXX}(1,3) \frac{\delta \Sigma_x(3,4) }{ \delta \rho(2)} G_\text{EXX}(4,1) \\
   + \int d3 d4 ~ \frac{\delta G_\text{EXX}(1,3) }{ \delta \rho(2) }
      \left[ \Sigma_x(3,4) - \delta(3,4) v_\text{EXX}(3) \right] G_\text{EXX} (4,1)  \\
   + \int d3 d4 ~ G_\text{EXX} (1,3)
      \left[ \Sigma_x(3,4) - \delta(3,4) v_\text{EXX}(3) \right] 
        \frac{\delta G_\text{EXX}(4,1) }{ \delta \rho(2) }  .
\end{multline}

The linearized exchange operator is simply $\Sigma_x(1,2)= i G_\text{EXX}(1,2) v(1,2)$.
Therefore, the only quantity needed to carry on the derivation is the derivative of 
$G_\text{EXX}$ with respect to $\rho$. It can be evaluated along the following lines,
using standard functional analysis relations, and introducing the total KS potential within EXX
$v_\text{KS}$,
\begin{eqnarray}
\frac{\delta G_\text{EXX}(1,2) }{ \delta \rho(3) } 
 & = & \int d4 ~ 
     \frac{\delta G_\text{EXX}(1,2) }{ \delta v_\text{KS}(4) } 
     \frac{\delta v_\text{KS}(4) } {\delta \rho(3)} \nonumber \\
 & = & -\int d4 d5 d6 ~ G_\text{EXX}(1,5)  G_\text{EXX}(6,2)
    \frac{\delta G_\text{EXX}^{-1}(5,6) }{ \delta v_\text{KS} (4) }
    \frac{\delta v_\text{KS}(4) } {\delta \rho(3)} \nonumber \\
 & = & \int d4 ~ G_\text{EXX}(1,4)  G_\text{EXX}(4,2) \chi_0^{\text{EXX} -1}(4,3)  ,
 \label{eq:deltaG_deltarho}
\end{eqnarray}
where the last line was obtained from the Dyson equation 
$G_\text{EXX}^{-1} = G_0^{-1} - v_\text{KS}$ ($G_0$ standing for the free-electron Green's function)
and from the definition $ \chi_0^{\text{EXX} -1} =\delta v_\text{KS} / \delta \rho$.

Finally, by inserting Eq.~(\ref{eq:deltaG_deltarho}) into Eq.~(\ref{eq:lin_ss_diff}),
by multiplying by $\chi_0^\text{EXX}(2,5)$,
and integrating over the variable 2,
we obtain
the central equation for the linearized TD-EXX kernel
$f^\text{EXX, lin} = \delta v_\text{EXX} / \delta \rho$:
\begin{multline}
\int d2 d3 ~ \chi_0^\text{EXX}(1,3)  f^\text{EXX,lin}(3,2) \chi_0^\text{EXX}(2,5) =
 \int d3 d4 ~ G_\text{EXX}(1,3) G_\text{EXX}(4,1) v(3,4) G_\text{EXX}(3,5) G_\text{EXX}(5,4) \\
  -i G_\text{EXX}(1,5) \int d3 d4 ~ G_\text{EXX}(5,3) 
      \left[ \Sigma_x(3,4) - \delta(3,4)v_\text{EXX}(3) \right] G_\text{EXX}(4,1)  \\
  -i G_\text{EXX}(5,1) \int d3 d4 ~ G_\text{EXX}(1,3)
      \left[ \Sigma_x(3,4) - \delta(3,4)v_\text{EXX}(3) \right] G_\text{EXX}(4,5) ,
\end{multline}
with $\chi_0^\text{EXX} = -i G_\text{EXX} G_\text{EXX}$.
The kernel $f^\text{EXX, lin}$ can be split into two pieces $f^{(1), \text{lin}}$ and $f^{(2), \text{lin}}$,
the definitions of which stand respectively in Eqs.~(\ref{eq:f1lin}) and~(\ref{eq:f2lin}).
This partition is natural when looking at the analytical form of the terms.
It is further physically-driven, since the term $f^{(2), \text{lin}}$ accounts
for electron-hole interaction and the term $f^{(1), \text{lin}}$ for the
quasiparticle shift. \cite{fabien,pankratov}

\section{Link to TD-EXX}
\label{appendix:kimgorling}

This appendix shows that the kernel obtained in the previous
appendix is precisely the EXX kernel of Kim and G\"orling \cite{gorling_tdexx_prb}.
For simplification, let us name
$T^{(1a), \text{lin}}$ the first term of $\chi_0^\text{EXX}f^{(1), \text{lin}}\chi_0^\text{EXX}$,
$T^{(1b), \text{lin}}$ the second one,
and $T^{(2), \text{lin}} = \chi_0^\text{EXX} f^{(2), \text{lin}}\chi_0^\text{EXX}$.

We are about to introduce the expression of $G_\text{EXX}$
in the previous terms in order to recover all the
sixteen terms of Kim and G\"orling's kernel.
The time-ordered EXX Green's function $G_\text{EXX}$ in frequency domain is
\begin{equation}
G_\text{EXX}(\br_1,\br_2,\omega) =
 \sum_{i} \frac{\phi_i(\br_1) \phi_i^*(\br_2)}
               {\omega - \epsilon_i - i\eta ( 2 f_i - 1)} ,
\end{equation}
where $\phi_i$ and $\epsilon_i$ are the EXX KS wavefunctions and
energies for index $i$ (that contains also the $\bk$ point information).
$f_i$ is 1 for occupied states and 0 for empty states.

\subsection{Evaluation of $T^{(2), \text{lin}}$}

Let us first proceed with the electron-hole interaction term $T^{(2), \text{lin}}$.
It reads, after Fourier transform to frequency domain,
\begin{multline}
T^{(2), \text{lin}}(\br_1,\br_5,\omega) =
\frac{2}{(2\pi)^2} \sum_{ijkl}
\phi_i(\br_1)\phi_j^*(\br_1)
 \int d \omega_1
  \frac{1}
       {(\omega+\omega_1-\epsilon_i)(\omega_1-\epsilon_j)} \\
 \times \langle ik |v| jl \rangle
\phi_k^*(\br_5)\phi_l(\br_5)
 \int d \omega_2
  \frac{1}
       {(\omega+\omega_2-\epsilon_k)(\omega_2-\epsilon_l)} ,
\end{multline}
as the products in time space become convolutions of frequencies.
The factor 2 accounts for spin degeneracy.
The usual Coulomb integrals
\begin{equation}
\langle ik |v| jl \rangle = \int d \br_1 d\br_2 
   \phi_i^*(\br_1) \phi_k(\br_1)
             \frac{1}{|\br_1 - \br_2|}
   \phi_j(\br_2) \phi_l^*(\br_2) 
\end{equation}
have been introduced
and the $\pm i\eta$ factors in the denominators are still present, but not explicitly written
(they are unchanged with respect to the definition of $G_\text{EXX}$).

The frequency integrals are now calculated by virtue of the residue theorem
on a path that encloses either the upper half-plane, or the lower half-plane.
Contributions with all poles in the same half-plane vanish.
Consequently, the frequency integrals are
\begin{equation}
\int d \omega_1   \frac{1}
       {[\omega+\omega_1-\epsilon_i- i \eta (2 f_i - 1)][\omega_1-\epsilon_j- i \eta (2 f_j - 1)]}
 = 2 \pi i \frac{f_j-f_i}{\omega-(\epsilon_i-\epsilon_j)
             + i\eta(f_j-f_i)}  .
\end{equation}

The $T^{(2), \text{lin}}$ term finally reads
\begin{equation}
T^{(2), \text{lin}}(\br_1,\br_5,\omega) =
- 2 \sum_{ijkl}
(f_j-f_i)
 \frac{\phi_i(\br_1)\phi_j^*(\br_1)}
 {\omega-(\epsilon_i-\epsilon_j)+ i\eta(f_j-f_i)}
  \langle ik |v| jl \rangle 
(f_l-f_k)
 \frac{\phi_k^*(\br_5)\phi_l(\br_5)}
 {\omega-(\epsilon_k-\epsilon_l)+ i\eta(f_l-f_k)} .
\end{equation}
This expression for $T^{(2), \text{lin}}$ is equal to the $H_X^1$ and $H_X^2$ terms
of Ref.~\onlinecite{gorling_tdexx_prb}, except that the convergence factors $i\eta$
are of opposite sign for antiresonant terms.
In fact, the present derivation, starting from time-ordered Green's functions,
yields time-ordered quantities,
whereas Kim and G\"orling's derivation considers causal quantities.
When used adequately, this difference is not relevant in practical applications.

\subsection{Evaluation of $T^{(1a), \text{lin}}$ and $T^{(1b), \text{lin}}$}

Let us now turn to the contribution $T^{(1a), \text{lin}}$ to the linearized TD-EXX kernel.
$\Sigma_x$ is a static approximation for the self-energy, hence
in the frequency domain, $T^{(1a), \text{lin}}$ reads
\begin{multline}
T^{(1a), \text{lin}}(\br_1,\br_5,\omega) =
-\frac{2i}{2\pi} \sum_{ijk} \int d \omega_1
% \left[
 \frac{\phi_i(\br_1) \phi_i^*(\br_5)}{\omega+\omega_1-\epsilon_i- i \eta (2 f_i - 1)} 
% \right]
\\
 \times
 \frac{\phi_j(\br_5)}{\omega_1-\epsilon_j - i \eta (2 f_j - 1)}
 \langle j| \Sigma_x-v_\text{EXX} |k \rangle
 \frac{\phi_k^*(\br_1)}{\omega_1-\epsilon_k - i \eta (2 f_k - 1)} .
\end{multline}

Performing the integration on $\omega_1$ thanks to the residue theorem
gives a vanishing contribution if the $i$, $j$, $k$ states are all
occupied or all empty.
There are 6 non-vanishing terms corresponding to the other cases.
We will exemplify three of them in the following.
The three remaining ones are analogous.

If $i$ and $j$ are occupied and $k$ is empty, let us close the path
of integration in the lower half-plane. The enclosed poles are the
$\epsilon_k-i\eta$ that yield the residues:
\begin{equation}
-2\pi i \frac{f_i f_j (1 - f_k)}
             {[\omega-(\epsilon_i-\epsilon_k)-i\eta][\epsilon_k-\epsilon_j]} .
\end{equation}
If $i$ and $k$ are occupied and $j$ is empty, closing the path analogously 
in the lower half-plane retains the poles $\epsilon_j-i\eta$ that give
the residues:
\begin{equation}
-2\pi i \frac{f_i (1 - f_j) f_k}
             {[\omega-(\epsilon_i-\epsilon_j)-i\eta][\epsilon_j-\epsilon_k]} .
\end{equation}
If $j$ et $k$ are occupied and $i$ is empty, this retains poles
located at $\epsilon_i-\omega-i\eta$ with residues:
\begin{equation}
-2\pi i \frac{ ( 1 - f_i) f_j f_k}
             {[\omega-(\epsilon_i-\epsilon_j)-i\eta]
              [\omega-(\epsilon_i-\epsilon_k)-i\eta]}  .
\end{equation}
The three other terms correspond to the case with 2 empty states and 1 occupied.
The path of integration will be closed in the upper half-plane,
in order to retain only the poles from the occupied states.

$T^{(1a), \text{lin}}$ finally gives rise to 6 terms.
$T^{(1b), \text{lin}}$ will also account for 6 analogous terms.
The sum of $T^{(1a), \text{lin}}$ and $T^{(1b), \text{lin}}$, if written explicitly, is exactly
the terms $H_X^3+H_X^4$ of Kim and G\"orling, except once again
that the convergence factors $i\eta$ are opposite for antiresonant transitions.

This appendix showed that the linearized Sham-Schl\"uter equation indeed
yields the same TD-DFT kernel, as the one obtained by Kim and G\"orling in
Ref.~\onlinecite{gorling_tdexx_prb}.

\bibliography{td-exx}

%\newpage
%\includegraphics[width=0.9\columnwidth]{si_hf}
%\newpage
%\includegraphics[width=0.9\columnwidth]{si-rho-G}
%\newpage
%\includegraphics[width=0.9\columnwidth]{epsm1q}
%\newpage
%\includegraphics[width=0.9\columnwidth]{si-cLDA}

\end{document}